# *1/f* noise near the free surface of a semiconductor

Tomasz Blachowicz

Institute of Physics, Silesian University of Technology

Department of Electron Technology

Krzywoustego 2, PL-44100 Gliwice, POLAND

This paper describes the so-called *1/f* noise generated within the depleted region below a free semiconducting surface. It was shown that there is a link between the *1/f* noise, which can be controlled by the band-bending, and the very narrow energy states ($E<<kT$) located on the top surface and vanishing into the bulk region. Also, it was shown that the noise can be lowered when the band-bending is reduced. A quantitative description of this reduction was provided. This study considered the n-type semiconductor.

# I. INTRODUCTION

The $1/f$ noise is a physical effect originating from different sources with characteristic fluctuations of the noise power density proportional to $f^{-\gamma}$, where $\gamma = 1.0 \pm 0.1$ [1]. There are two main types of distinctions in noise analysis. The distinction between the number of occupation fluctuations and mobility fluctuations [2], and the distinction between bulk contributions and surface effects [3]. In addition, noise depends on other factors, such as: growing conditions, number of structural defects, surface quality, and chemical treatment like a passivation to name a few.

In applications, the $1/f$ noise has a very fundamental importance in low-power and microwave devices. Among many of them we can mention: AlGaN/GaN high electron mobility transistors (HEMTs) where the device characteristics degrade as a result of trapping of electrons in the active surface area [4, 5], InGaAsP/InP heterostructures used for laser diodes [6], ambipolar semiconductors (Ge) [7], Al-GaAs Schottky barriers [8], Si N-MOSFETs [9], MESFETS [10], and silicon quantum wires [11].

In terms of the general picture of the *1/f* noise, the trap states created by surface defects and the trap states distributed in the undersurface atomic layers should be considered as a starting point for the analysis. Then, the trap states existence is equivalent to a relaxation-time distribution of electric carriers needed to explain of the noise origin. Thus, the energy distribution of trap concentrations per se is an important factor to be taken into the account.

The good example of efficiently working theory of the bulk $1/f$ noise, involving time-energy and concentration-energy distributions can be found in [12] where the tail states in highly-doped semiconductors and the resulting conduction-band bending were considered. This report, however, illustrates how the low-frequency $1/f$ noise in a depleted region of a semiconductor is induced by surface trap states.

Depleted region is associated with the natural band-bending and the existence of many undersurface defects which contribute to the relaxation time distribution of electric carriers. Also, it is discussed here what the efficiency of noise reduction is and how it can be caused by the removal of band-bending. In addition, the clear, quantitative measure of the $1/f$ noise reduction due to passivation related to the depleted region of semiconductor was introduced. It should be mentioned that reducing noise intensity sometimes is not practical, and many devices utilize the band bending such as, for example, Schottky diodes [13, 14].

## II. THE NOISE vs. TIME-RELAXATION DISTRIBUTIONS

The starting postulate for every noise analysis is the existence of the relaxation-time distribution $g(\tau)$ along with the range of the time constants $\langle \tau_1, \tau_2 \rangle$, for example

$$g(\tau) \stackrel{def}{=} \frac{1}{\ln(\tau_2/\tau_1)} \frac{1}{\tau} \qquad (2\pi/\tau_2) << \omega << (2\pi/\tau_1). \tag{1}$$

From the above results the formal definition of the frequency noise of the *1/f* type ($1/f = 2\pi/\omega$), expressed in (1/Hz) units, is given by the following integral [1]

$$S_\omega(f) \stackrel{def}{=} \overline{\Delta N^2} \int_{\tau_1}^{\tau_2} g(\tau) \frac{4\tau}{1+\omega^2\tau^2} d\tau = \overline{\Delta N^2} \frac{1}{\ln(\tau_2/\tau_1)} \frac{1}{f}, \tag{2}$$

where $\overline{\Delta N^2}$ is the mean-square value of the fluctuation in the number of carriers.

In relation to the depletion region located below the free semiconductor surface, we can postulate that the trap's time-life distribution follows the band bending with the depth *x* measured from the top surface, or

$$\tau(x) = \tau(0)\exp(-x/w), \tag{3}$$

where $\tau(0)$ is the relaxation time at the free surface and *w* is the depletion layer width. This postulate is quite arbitrary and doesn't take into an account any specific detrapping mechanism, like tunneling or thermal activation. From the above, Eq. 3, we can derive the relaxation-time distribution which equals

$$g(x) = \frac{dN_t}{d\tau} = \frac{dN_t}{dx} \cdot \frac{dx}{d\tau} = f(x) \cdot \frac{w}{\tau(x)} \approx \frac{1}{\tau(x)}, \tag{4}$$

where $N_t(x)$ is the trap concentration distribution function inside the depletion layer. The $N_t(x)$ function, in general, can be quite arbitrary. However, for the special case of trap states induced on semiconductor surface, and consequently, in the undersurface depleted region, we can assume that the $N_t(x)$ distribution equals $N_t(x) = N_{tx}(0)\exp(-x/w) \cong N_{tx}(0)(1-x/w)$. This does not change the general result of Eq. 4, since the time-life distribution is still inversely proportional to the time-constant $\tau$. Thus, we have

$$g(\tau) = \frac{dN_t}{d\tau} = \frac{dN_t}{dx} \cdot \frac{dx}{d\tau} = \left(-\frac{N_{tx}(0)}{w}\right) \cdot \left(\frac{w}{\tau(x)}\right) \approx \frac{const}{\tau(x)} \quad . \tag{5}$$

Eqs. 4-5 warrant the existence of noise (comp. Eqs. 1-2), however, as it will be shown below, these equations do not warrant the existence of the *1/f* noise. Again, Eqs. 4-5 represent a physical system which lacks thermodynamic equilibrium within the relaxation-time scale, as surface states can interact with free carriers in the undersurface region contributing to the noise effects.

The starting points for calculation in the energy scale are the two dependencies for the time-relaxation distribution $\tau(E)$ and the number of trap states distribution $N_t(E)$, namely

$$\tau(E) = \tau(E_c)F(E_c)\exp[-(E-E_c)/E_\tau], \quad E_\tau = E_\tau(w, kT) , \tag{6}$$

$$N_t(E) = N_{tE}(E_c)\exp[-(E-E_c)/E_n], \quad E_n = E_n(w, kT) \approx kT . \tag{7}$$

Both distributions are extended over the bottom edge of the conduction-band bulk-level $E_c$ (Fig. 1), their widths are represented by the $E_\tau$ and $E_n$ parameters, for the time-relaxation and the traps density, respectively. $F(E_c)$ is the value of Fermi-Dirac function at the $E_c$ level and the values of $\tau(E_c)$ and $N_{tE}(E_c)$ are obvious at the $E_c$ level. The energy distribution of the trap concentration given in equation (7) with $E_n \approx kT$ can results in white noise not 1/f noise in some cases [15].

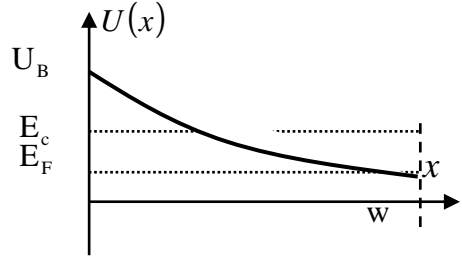

Fig. 1. The band structure U(x) used in calculations. Descriptions: $E_F$ – the Fermi level, $U_B$ - the band-bending, $E_c$ – the bulk level of conduction band, w – the depleted region width.

Next, the infinitesimal contribution to noise (comp. Eq. 2) induced at the energy level $E$ by the trap concentration $N_t$, for the volume sample $V$, with $F$ - the probability of absorption (catching of a carrier) or the re-emission from a trap level with the probability $(1-F)$, equals

$$dS_\omega(E) = \frac{N_t(E)}{V} F(1-F) \frac{4\tau(E)}{1+\omega^2\tau^2(E)} dE, \qquad (8)$$

while the Fermi-Dirac distribution function [16]

$$F = \frac{1}{1+\exp[-(E_F-E)/kT]} \cong \exp[(E_F-E)/kT] \qquad (9)$$

can be simplified for the assumed non-degenerate case. This is why the integral of Eq. 8, calculated around the $E_c$ level, in the $\langle E_c - E_n; E_c + U_B \rangle$ range up to the band-bending $U_B$, equals

$$S_\omega(E) \cong D \int_{E_c-E_n}^{E_c+U_B} \frac{\exp[-b(E-E_c)]}{1+a^2 \exp[-c(E-E_c)]} dE, \qquad (10)$$

where some simplifying symbols were introduced. These are as follows

$$a = \omega\tau(E_c)F(E_c), \; b = \frac{1}{E_n} + \frac{1}{E_\tau}, \; c = \frac{2}{E_\tau}, \; D = \frac{4N_{tE}(E_c)\tau(E_c)F^2(E_c)(1-F(E_c))}{V}. \qquad (11)$$

Eq. 10 can be solved for several assumptions. Thus, if the widths $E_\tau$ and $E_n$ are comparable then the integral is equal to

$$S_\omega(E_\tau \approx E_n) \cong \frac{D E_\tau}{2\tau^2(E_c) F^2(E_c)} \ln\left(\frac{1+a^2 \exp(2E_n/E_\tau)}{1+a^2 \exp(-2U_B/E_\tau)}\right) \cdot \frac{1}{\omega^2}. \qquad (12)$$

and we obtain $1/f^2$ noise which is not the case we are looking for. Therefore, if we were to assume that the relaxation time distribution of traps is very narrow, we get the following relationship

$$S_\omega(E_\tau \ll E_n) \cong \frac{D E_\tau}{\tau(E_c) F(E_c)} \underbrace{\left[\arctg\left(a e^{E_n/E_\tau}\right) - \arctg\left(a e^{-U_B/E_\tau}\right)\right]}_{\cong \pi \; (E_\tau \ll E_n, \; E_\tau \ll U_B)} \cdot \frac{1}{\omega}, \qquad (13)$$

the $1/f$ nonlinear function of several energy and time parameters, thus, the function of the band bending $U_B$. From the physical point of view the condition $E_\tau \ll E_n$ describes non-correlated trap states. In other words traps are independent, out of memory, of the catch and re-emit type.

Finally, it is very interesting to show the following limiting cases. The first one is related to the lack of time-relaxation distribution. Thus, if $E_\tau \to 0$ in Eq. 13, then the noise goes to zero $S_\omega \to 0$. The second one is related to the reduction of band bending, possibly by the use of passivation method. Namely, for the non-correlated trap states, $E_\tau \ll E_n$, but additionally, assuming the similar inequality for the band bending, $E_\tau \ll U_B$, we obtain

$$S_\omega(E_\tau \ll E_n, E_\tau \ll U_B) \cong \frac{4 N_{tE}(E_c) E_\tau}{V} F(E_c)(1 - F(E_c)) \frac{\pi}{\omega}, \qquad (14)$$

Surprisingly, if we were to additionally apply the $U_B = 0$ limit we obtain the noise which is reduced by a factor of 2, that is

$$S_\omega(E_\tau \ll E_n, U_B = 0) \cong \frac{4 N_{tE}(E_c) E_\tau}{V} F(E_c)(1 - F(E_c)) \frac{\pi}{2\omega}. \qquad (15)$$

The last two relations express the general physical meaning of the possible noise reduction in the depletion layer of a semiconductor. For the very narrow, non-correlated trap states assumed above, the $1/f$ noise can be reduced by half if the band bending reaches theoretical zero value.

## III. CONCLUSIONS

The band bending and the existence of the depletion layer are the two counterparting effects resulting from the natural breakdown of periodic atomic bonds at the free surface of a semiconductor. Consequently, the surface concentration of the electric carriers is different from that of the bulk region, and also, the undersurface region is depleted in a sense, that the distribution of carriers changes continuously with the distance measured from a surface.

Thus, we deal with a physical system with distributed in-depth properties and the distributed relaxation times of trap states. Together, they create the basis for the existence of the *1/f* noise. It seems that this conclusion is quite general and can be applied for all situations where there is band-bending, not only those caused by broken bonds at the surface.

It was shown that the reduction of band bending can improve $1/f$ characteristics. However, even if the band bending can be removed, the noise can not be completely neglected at temperatures different from the absolute zero and for realistic materials with surface and undersurface defects. It should be also remembered that the band bending in specific practical solutions is a needed effect supporting performance of devices.

[1] F. N. Hooge, IEEE Trans. Electron. Dev. 41, 1926 (1994).

[2] A. T. Hatzopoulos, N. Arpatzanis, D. H. Tassis, C. A Dimitriadis, F. Templier, and M. Oudwan, G. Kamarinos, Solid-State Electron. 51, 726 (2007).

[3] P. Dutta and P. M. Horn, Rev. Mod. Phys. 53, 497 (1981).

[4] W. Lu, V. Kumar, R. Schwindt, E. Piner, and I. Adesida, Solid-State Electron. 46, 1441 (2002).

[5] S. A. Vitusevich, M. V. Petrychuk, S. V. Danylyuk, A. M. Kuraki, N. Klein, and A. E. Belyaev, Phys. Stat. Sol. (a) 202, 816 (2005)..

[6] R. Hakimi R and M. C. Amann, Semicond. Sci. Technol. 12, 778 (1997).

[7] T. Dilmi, A. Chovet, and P. Viktorovitch, J. Appl. Phys. 50, 5348 (1979).

[8] S. Meškinis, G. Balčaitis, J. Matukas, and V. Palenskis, Solid-State Electron. 47, 1713 (2003).

[9] L. K. J. Vandamme, X. Li, and D. Rigaud, IEEE Trans. Electron. Dev. 41, 1936 (1994).

[10] M. Chertouk and A. Chovet, IEEE Trans. Electron. Dev. 43, 123 (1996).

[11] A. Balandin, K. L. Wang, A. Svizhenko, and S. Bandyopadhyay, IEEE Trans. Electron. Dev. 46, 1240 (1999).

[12] E. Borovitskaya and M. S. Shur, Solid-State Electron. 45, 1067 (2001).

[13] S. T. Hsu, IEEE Trans. Electron. Dev. 17, 496 (1970).

[14] S. T. Hsu, IEEE Trans. Electron. Dev.18, 882 (1971).

[15] J. I. Lee, J. Brini, A. Chovet, and C. A. Dimitriadis, Solid-State Electro. 43, 2181 (1999).

[16] W. Shockley and W. T. Read, Phys. Rev. 87, 835 (1950).